\documentclass{jpsj2}

\def\runauthor{Online-Journal Subcommittee}

\title{%
Vortices in $p$-Wave Superfluids of Trapped Fermionic Atom Gases
}

\author{%
Yasumasa Tsutsumi\thanks{E-mail address: tsutsumi@mp.okayama-u.ac.jp} and Kazushige Machida
}

\inst{%
Department of Physics, Okayama University,
Okayama 700-8530
}

\recdate{\today}

\abst{%
In order to help detecting superfluidity,
we theoretically investigate $p$-wave pairing superfluids in neutral Fermion atom gases
confined by a three dimensimentional (3D) harmonic potential.
The Ginzburg-Landau framework, which is generic for $p$-wave superfluids, is used to describe the
order parameter spatial structure, or texture characterized by the $l$-vector both at rest and under rotation.
The $l$-vector configuration is strongly contrained by the boundary condition due to a trap.
It is found that the ground state textures exhibit spontaneous supercurrent at rest both
cigar and pancake shape traps. The current direction depends on the trapping shape.
Under rotation a pair of half-quantum vortex with half-winding number enters a system and is stabilized
for both trap geometries. We give detailed explanation for their 3D structure.
The deformations of the condensate shape are seen with increasing the rotation speed,
which is tightly  connected with the underlying vortex formation where the condensates are 
depressed in the vortex core.
}

\kword{
$p$-wave pairing, superfluids, neutral Fermion atom gases, texture, vortex
}

\begin{document}
\maketitle

\section{Introduction}

Superfluids with multi-component order parameter (OP) are omni-present,
exemplified by spinor Bose-Einstein condensates (BEC)\cite{dan,stenger,ohmi,ho} 
in cold Bosonic atom gases with $^{23}$Na or $^{87}$Rb, strongly interacting 
Fermionic liquid $^3$He atoms\cite{leggett,wolfle,volovik,fetter,salomaa}, and certain
heavy Fermion superconductors, such as UPt$_3$\cite{machida,sauls} or
color superconductivity in dense 
quark-gluon plasmas in high energy physics\cite{buballa,alford}.
This branch of physics is of great interest from a fundamental physics view point as it is expected 
to exhibit rich topological defect structures or vortices and allow
investigation of a new phase of matter.
Those vortices that can accommodate the Majorana zero mode at 
the core may be useful in quantum computing\cite{sarma}.
We will show in this paper that half-quantum vortices (HQV) are stabilized,
which is a candidate to lead to the Majorana particles in the core\cite{sarma}.
Therefore the present $p$-wave superfluids in general provide a tesing ground to
explore rich physics associated with topological defects. 
Previously we have demonstrated that superfluid $^3$He-A
phase is another fertile ground to find the Majorana particle\cite{tsutsumi,kawakami}.
This is the second example to  show HQV
as a stable vortex in our study\cite{kawakami} where superfluid $^3$He-A
phase confined between parallel plates is demonstrated to exhibit HQV.

Recently, $p$-wave resonance superfluidity attracts  much attention in Fermionic alkaline atom gases, 
such as $^6$Li\cite{zheng,schunck,inada} and $^{40}$K\cite{jin1,jin2,jin3},
both experimentally and theoretically\cite{yip,leo,ho2,quader,ohashi}.
Experiments into achieving $p$-wave resonance superfluidity are steadily 
progressing\cite{inada} and hence it is timely and necessary to 
consider the generic properties of $p$-wave superfluidity both at rest and 
under rotation to detect its superfluidity,
which is signified by non-trivial topological structures or vortices. 
In this respect, the interest is in the spatial structure,
i.e. texture, of the nine component OPs describing the spin triplet $p$-wave superfluidity.
A $p$-wave Feshbach resonance occurs at the different magnetic field
in each hyperfine spin state of Cooper pair.
Since the spin state of superfluidity is fixed by the external magnetic field,
the spin  degrees of freedom are frozen,
and hence only the orbital degrees of freedom are active.
The order parameter space consisting of orbital three components
is analogous to that of 
superfluid $^3$He, in particular, 
the A phase to which spin and orbital state can be separated\cite{wolfle,leggett,volovik,fetter},
where the OP  is described by a tensor 
\begin{equation}
A_{\mu i}=d_{\mu}A_i  \ (\mu, i=x,y,z). 
\end{equation}
The $d_{\mu}$ and $A_i$ describe the spin and orbital
state of a Cooper pair, respectively. 
Here the OP is characterized only by $A_i$,
described as 
\begin{equation}
\Delta (\hat{\bf p})=A_x\hat{p}_x+A_y\hat{p}_y+A_z\hat{p}_z.
\end{equation}
In a sense, a $p$-wave superfluid is analogous to the ``spinless'' superfluid $^3$He-A phase\cite{tsutsumi}.

The dipole-dipole interaction between two alkaline atoms
acts to split the relative orbital state for two particles, depending on 
the projections of the orbital angular momentum, either $m_l=\pm 1$ or $m_l=0$.
This results in breaking of the degeneracy between 
\begin{equation}
\hat{p}_{\pm}=\mp {1\over \sqrt{2}}(\hat{p}_x\pm i\hat{p}_y) \qquad {\rm and} \qquad \hat{p}_0=\hat{p}_z.
\end{equation}
This splitting was estimated to be large for $^{40}$K by Cheng and Yip\cite{yip},
evidenced by the clear difference in the Feshbach resonance
magnetic fields (splitting field = 0.47$\pm$0.08G)\cite{jin2}.
For $^6$Li, the splitting may be small, as an experiment conducted in a magnetic field of $H=158.5(7) $G 
shows no clear resonance splitting\cite{inada}.

A critical difference between superfluid $^3$He and a $p$-wave resonance superfluid of 
atom gases lies in the boundary conditions. 
In the superfluid $^3$He-A phase the $l$-vector, 
which is the orbital angular momentum of Cooper pairs, is always perpendicular to a
hard wall so that the perpendicular particle motion is suppressed.
In other words, the point nodes in the
$l$-vector direction touch the hard wall so as to minimize the condensation energy loss 
at the boundary\cite{wolfle}.
On the other hand,
atom gases are confined by a three dimensional (3D) harmonic trap potential,
where the condensation energy density 
gradually decreases towards the outer region, the $l$-vector tends to 
align parallel to the circumference. This orientation is advantageous because 
the condensation energy is maximally gained by allowing the point nodes to move out from the system.
The trap potential is easily controlled, resulting in various shapes, such as
cigar or pancake shapes. As we describe below, the trapping potential can be
an important tool to control the 3D texture.
Indeed, the 3D trapping structure constrains the possible textures. 
Our purpose is to investigate a possible 3D textures
in a 3D harmonic trap potential and thereby to help identifying $p$-wave superfluidity.

The organization of this paper is as follows: 
We employ the Ginzburg-Landau (GL) framework which relies only on
global symmetry principle.
The GL free energy functional form is introduced and the relevant physical quantities,
such as supercurrent and $l$-vector are given in \textsection{} 2.
We show the phase diagram of the stable states for uniform and infinite system,
and explain how to numerically examine realistic confined systems in \textsection{} 3.
Section 4 presents the stable texture in the cigar shape trap,
especially, we mention the spontaneous supercurrent at rest 
and the half-quantum vortex (HQV) under rotation.
In \textsection{} 5, we show the different textures stabilized in the pancake shape.
The final \textsection{} 6 is devoted to a summary and discussion.
A short version of the present paper is to be published\cite{tsutsumi2}.

\section{Formulation}

Here we employ the GL framework\cite{wolfle}. This framework is general and flexible enough to allow us to
examine a generic topological structure,
and applicable to cold Fermionic atom gases with a harmonic trap potential under $k_BT_c \gg \hbar \omega$,
where $T_c$ and $\omega$ are the transition temperature and the trap frequency, respectively\cite{baranov}.
In terms of the tensor $A_{\mu i}$ forming OP of $p$-wave pairing the most general GL functional density $f_{\rm bulk}$ 
for the bulk condensation energy up to fourth order is described as 
\begin{align}
f_{\rm bulk}=-\alpha_i A^*_{\mu i}A_{\mu i}+\beta_1A^*_{\mu i}A^*_{\mu i}A_{\nu j}A_{\nu j}
+\beta_2 A^*_{\mu i}A_{\mu i}A^*_{\nu j}A_{\nu j}
+&\beta_3 A^*_{\mu i}A^*_{\nu i}A_{\mu j}A_{\nu j}\nonumber \\
+\beta_4 A^*_{\mu i}A_{\nu i}A^*_{\nu j}A_{\mu j}
+&\beta_5 A^*_{\mu i}A_{\nu i}A_{\nu j}A^*_{\mu j},
\end{align}
which is invariant under spin and real space rotations 
in addition to the gauge invariance U(1)$\times$SO$^{(S)}$(3)$\times$SO$^{(L)}$(3).
The fourth order terms are characterized by five independent
invariants, $\beta_1\sim\beta_5$ in general\cite{wolfle}.
Since the spin degrees of freedom are frozen due to applied magnetic field
for magnetic Feshbach resonance,
only the orbital degrees of freedom $A_i$ in $ A_{\mu i}= d_{\mu}A_i$  are active. Namely it reduces to 
\begin{equation}
f_{\rm bulk}=-\alpha_0 (1-t_i) A_i^* A_i + \beta_{24} A_i^* A_i A_j^* A_j + \beta_3 A_i^* A_i^* A_j A_j ,
\label{fbulk}
\end{equation}
where $\beta_{24}=\beta_2+\beta_4$ and  $t_i=T/T_{ci}$ ($T_{ci}$ is the transition temperature for the $i$-component).
As mentioned, the dipole-dipole interaction causes splitting of the transition 
temperatures into two groups $T_{cx}=T_{cy}$ and $T_{cz}$.
We introduce $\alpha=T_{cx}/T_{cz}$, which indicates the degree of the broken symmetry of the system
and characterizes atomic species used\cite{jin2}.
The pairing state having the orbital projection $m_l=0$ is favorable over $m_l=\pm 1$, namely $0<\alpha <1$ 
due to the dipole-dipole interaction\cite{jin2}.
The three components become degenerate for $\alpha\rightarrow 1$. When $\alpha\rightarrow 0$, 
the polar state  with the OP $\Delta(\hat{\bf p})=A_z \hat{p}_z$ tends to be stable.

The gradient energy consisting of the three independent terms\cite{wolfle} is given by 
\begin{eqnarray}
f_{\rm grad}=K_1(\partial_i^* A_j^* )(\partial_i A_j )+K_2(\partial_i^* A_j^* )(\partial_j A_i )
+K_3(\partial_i^* A_i^* )(\partial_j A_j ) .
\label{fgrad}
\end{eqnarray}
The centrifugal potential energy due to rotation with ${\bf \Omega}$, which is derived in Appendix A,  is written as 
\begin{equation}
f_{\rm cent}=-{m^2\over \hbar^2}\Omega^2\rho^2(K_1A_i^* A_i + K_2 |A_{\theta}|^2 + K_3|A_{\theta}|^2 ),
\label{fc}
\end{equation}
where $\partial_i=\nabla_i-i(m / \hbar)({\bf \Omega \times r})_i $.
For ${\bf \Omega \parallel \hat{z}}$,
$\rho^2=x^2+y^2$ and $A_{\theta}=-A_x \sin{\theta} + A_y \cos{\theta}$ in the cylindrical coordinates.

The GL parameters $\alpha_0$, $\beta_{24}=\beta_2+\beta_4$, $\beta_3$ and $K_1=K_2=K_3=K$ are 
estimated  by taking the weak coupling approximation, assuming the
Fermi sphere\cite{wolfle}:
\begin{equation}
\alpha_0={N(0)\over 3}, \ \beta_2=\beta_3=\beta_4={7\zeta(3)N(0)\over 120(\pi k_BT_c)^2} \equiv \beta
\end{equation}
 and 
\begin{equation}
K={7\zeta(3)N(0)(\hbar v_F)^2\over240(\pi k_BT_c)^2}
\end{equation}
where $N(0)$ is the density of states at the Fermi level
and $v_F$ is the Fermi velocity.
The weak coupling approximation should be a good guide for understanding the generic properties of the 
$p$-wave superfluids of atom gases because it has been applied successfully, even to liquid $^3$He with strong interacting
Fermions and only small additional strong corrections\cite{wolfle}.

It is convenient to discuss \eqref{fbulk}, \eqref{fgrad} and \eqref{fc} in the following  dimensionless units,
\begin{equation}
{A_i\over A^{(0)}} \rightarrow A_i, \  {r\over\xi_0} \rightarrow r, \ {\Omega \over\Omega^{(0)} }\rightarrow \Omega
\end{equation}
with the zero-temperature GL coherence length 
\begin{equation}
\xi_0=\sqrt{K\over\alpha_0 }
={\sqrt{7\zeta (3)\over 80\pi^2}}{\hbar v_F\over k_BT_c}.
\end{equation}
The units of OP  and the angular velocity are
\begin{equation}
A^{(0)}={\sqrt{\alpha_0\over 2\beta}}={\sqrt{20\pi^2\over 7\zeta (3)}} k_BT_c,
\end{equation}
\begin{equation}
\Omega^{(0)}={\hbar \over m}{1\over \xi_0^2}.
\end{equation}
In the dimensionless unit, \eqref{fbulk}, \eqref{fgrad} and \eqref{fc} are written as
\begin{align}
f_{\rm bulk}=&-(1-t_i) A_i^* A_i + A_i^* A_i A_j^* A_j + \frac{1}{2}A_i^* A_i^* A_j A_j , 
\label{fbulk2}\\
f_{\rm grad}=&(\partial_i^* A_j^* )(\partial_i A_j )+(\partial_i^* A_j^* )(\partial_j A_i )
+(\partial_i^* A_i^* )(\partial_j A_j ), 
\label{fgrad2}\\
f_{\rm cent}=&-\Omega^2\rho^2(A_i^* A_i + 2|A_{\theta}|^2 ),
\label{fc2}
\end{align}
respectively, where $\partial_i=\nabla_i-i({\bf \Omega \times r})_i $.
From now on we use the dimensionless expressions.

The harmonic trap potential term\cite{baranov} is
\begin{equation}
f_{\rm harmonic}=\omega_{\perp}^2(\rho^2+\lambda^2z^2)A_i^* A_i ,
\end{equation}
where the dimensionless radial confining potential is $\omega_{\perp}$ and
the anisotropy of the harmonic trap is expressed as $\lambda\equiv \omega_{z}/\omega_{\perp}$. 
The harmonic trap potential term acts to lower the transition temperatures.
It is interesting to note that the centrifugal 
potential leads to the non-trivial form,
because the OP label implies the orbital angular momentum, a feature absent in a spinor BEC\cite{mizushima}.
The extra factor of $2|A_{\theta}|^2$ in the above form of \eqref{fc2} becomes important
when evaluating the critical angular velocity $\Omega_{\rm cr}$, above which the superfluid flies apart.
That is, $\Omega_{\rm cr}=\omega_{\perp}/\sqrt 3$
is greatly reduced from the usual case ($\Omega_{\rm cr}=\omega_{\perp}$).

The relevant physical quantities are described in terms of the OP $A_i({\bf r})$ as follows.
The total free energy:
\begin{equation}
F = \int d^3r \left( f_{\rm bulk}  + f_{\rm grad} + f_{\rm cent} + f_{\rm harmonic}\right) .
\label{F}
\end{equation}
The current density:
\begin{equation}
j_i({\bf r}) \equiv 2 {\rm Im} \left[ A_j^*\nabla_i A_j + A_j^*\nabla_j A_i + A_i^*\nabla_j A_j \right] .
\end{equation}
The $l$-vector:
\begin{equation}
l_i({\bf r}) \equiv -i \epsilon_{ijk} {A_j^* A_k \over \mid \Delta ({\bf r}) \mid^2} 
\end{equation}
with
\begin{equation}
\mid \Delta ({\bf r}) \mid^2=A_i^* A_i .
\end{equation}

\section{Preriminary Considerations}

Before considering a realistic confined system, we first investigate an infinite system at rest.
Then the state minimizing the condensation energy \eqref{fbulk2} is realized.
As shown in Appendix B, the phase diagram in Fig. \ref{f1} consists of
the three phases, A, B, and normal (N) phases.
The B phase is described by $\Delta(\hat{\bf p}) = A_z \hat{p}_z$, i.e. the polar state.
The A phase is described by a chiral OP expressed by
\begin{eqnarray}
\Delta(\hat{\bf p}) = A_z (\hat{p}_z+i\gamma \hat{p}_{\perp}),
\end{eqnarray}
where
$( 0 < \gamma \le 1)$ with $\hat{p}_{\perp}=\hat{p}_x\cos \phi + \hat{p}_y \sin \phi$
($\phi$ is arbitrary). This phase  breaks the time reversal symmetry.
The value of $\gamma$ depends on temperature and anisotropy $\alpha$,
\begin{eqnarray}
\gamma = \frac{2-(3-\alpha )t_x}{2+(1-3\alpha )t_x}. 
\end{eqnarray}
In the absence of the dipole-dipole interaction ($\alpha =1$), the value of $\gamma$ is unity.
As the dipole-dipole interaction increases ($\alpha$ decreases), $\gamma$ decreases.
The decrease of $\gamma$ becomes larger at high temperature.
When the value of $\gamma$ vanishes, the second order transition from the A phase to the B phase takes place. 
The transition temperature $t_c$ is given by
\begin{eqnarray}
t_c=\frac{2}{3-\alpha}.
\end{eqnarray}
The phase diagram corresponds to the case being not so large Feshbach resonance splitting in BCS regime\cite{leo}.
However, it is warned that the determined phase boundary is of qualitative at low temperatures
which is beyond the GL framework.
In the following we examine the A phase in confined geometries.

In order to obtain stable texture of the condensates in a realistic harmonic potential, 
we have identified stationary solutions by numerically solving the variational
equations: $\delta f({\bf r})/\delta A_i({\bf r})=0$ in three dimensions
where $f({\bf r})$ is the GL energy density functional, the integrand of \eqref{F}.
We start with various initial configurations,
including singular vortex state and non-singular vortex state, 
and determine the most stable texture by comparing the total GL energy \eqref{F}.

\begin{figure}
\begin{center}
\includegraphics[width=6cm]{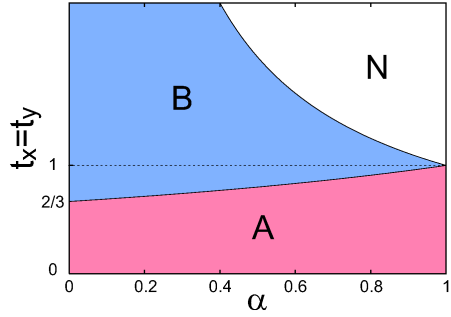}
\end{center}
\caption{(Color online) 
Phase diagram of the $p$-wave pairing state in an infinite system at rest,
showing temperature ($t_x=t_y$) versus anisotropy $\alpha\equiv T_{cx}/T_{cz}$.
N: normal state, A: chiral state $A_z (\hat{p}_z+i\gamma \hat{p}_{\perp})$ 
and B: polar state $A_z \hat{p}_z$. 
}
\label{f1}
\end{figure}

\section{Cigar Shape Trap}

We first consider the stable texture for a cigar shape trap with the trap anisotropy $\lambda =0.2$.
We take $80 \times 80 \times 120$ meshes with the cloud sizes for the Thomas-Fermi approximation 
$R_x=R_y=10$ and $R_z=50$.
We fix the temperature at $t_x=t_y=0.4$ and the anisotropy parameter $\alpha=0.9$.

\subsection{ Stable texture at rest}

\begin{figure}
\begin{center}
\includegraphics[width=8cm]{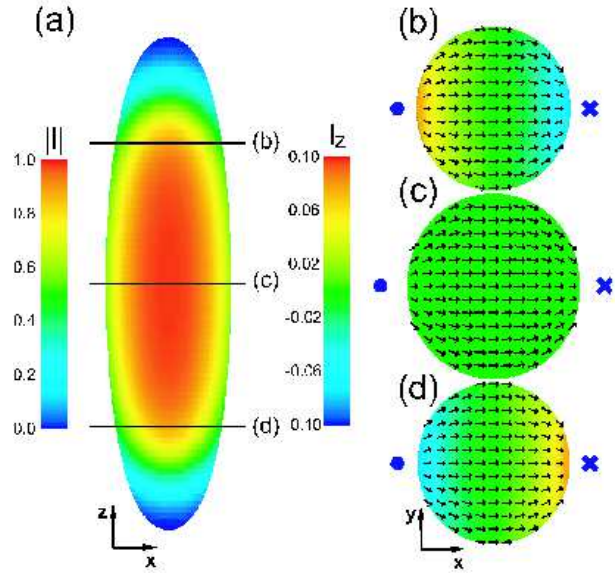}
\end{center}
\caption{(Color online) 
Stable texture at rest for cigar trap $\lambda=0.2$.
(a) Distribution of $|{\bf l}|$ in the $z$-$x$ plane. (b)-(d) Three cross sections indicated in (a),
showing $l_z$  (color) and $l_x$ and $l_y$ components (arrows),
where the dot and cross marks indicate the imaginative source and sink for the
$l$-vector stream lines.
}
\label{f2}
\end{figure}

The stable $l$-vector texture at rest is shown in Fig. \ref{f2}.
Figure \ref{f2}(a) displays the amplitude distribution of the $l$-vectors.
It can be seen that  the amplitude $|{\bf l} |$ is maximum in the central region, and towards the outer regions $|{\bf l} | $
decreases gradually.
At the top and bottom ends, the polar state is realized where the $l$-vector vanishes.
Three cross sections are shown in Figs. \ref{f2}(b)-(d).
In Fig. \ref{f2}(c), which corresponds to the middle cross section, the $l$-vectors lie in 
the $x$-$y$ plane, showing  a streamline type pattern in which the $l$-vectors
follow the circumference, like a fluid streaming along a circular boundary.  Outside the condensates, 
an unseen sink and source of the $l$-vectors exist, giving two imaginary focal points
situated outside. Namely the left dot and right cross marks in Figs. \ref{f2}(b)-(d) correspond to the source and sink
where the $l$-vectors appear and disappear. 
This streamline like texture contrasts with the so-called Pan-Am texture 
in superfluid $^3$He-A phase\cite{wolfle} where the $l$-vectors tend to point
perpendicular to the wall due to the boundary condition.
In the upper (Fig. \ref{f2}(b)) and lower (Fig. \ref{f2}(d))
cross sections the streamline texture is maintained, but an $l_z$ component appears in addition.

\begin{figure}
\begin{center}
\includegraphics[width=9cm]{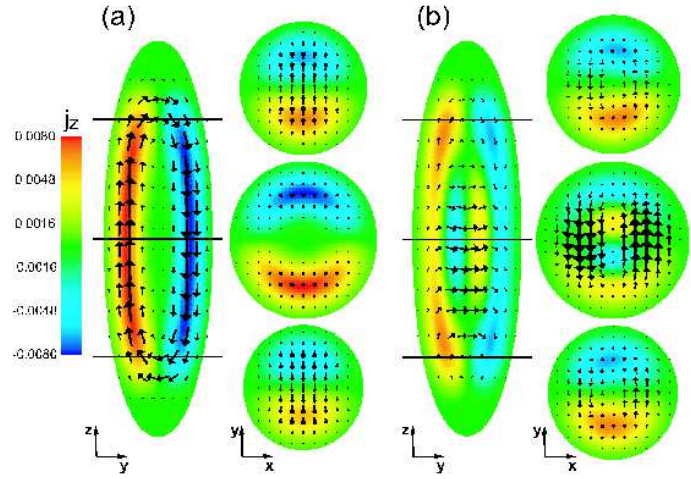}
\end{center}
\caption{(Color online) 
(a) Spontaneous circulating current flows at rest ($\Omega=0$)  in the $z$-$y$ plane along the $z$ direction.
Three cross sections indicated in the left figure,
showing $j_z$  (color) and $j_x$ and $j_y$ components (arrows).
(b)The current pattern under rotation ($\Omega=0.3$). It is seen that the in-plane components
$j_x$ and $j_y$ increase, producing the in-plane circular current due to rotation.
}
\label{f3}
\end{figure}

The associated supercurrent structure is depicted in Fig. \ref{f3}.
The $j_z$ component shows a circulation supercurrent along the $z$ axis (see Fig. \ref{f3}(a)).
Since $\Omega=0$, this circulation supercurrent is spontaneously generated.
This non-trivial condensates flow can be explained in supercurrent characteristic of chiral $p$-wave superfluid:
\begin{eqnarray}
{\bf j} = \mbox{\boldmath $\rho$}_s {\bf v}_s + {\bf C} ( \nabla\times {\bf l}),
\end{eqnarray}
where $\mbox{\boldmath $\rho$}_s$ and ${\bf C}$ are diagonal tensor coefficients.
The supercurrent has a structure that is similar to that for the total charge-current density 
of a system in classical electrodynamics.
The curent in classical electrodynamics is composed of two parts, 
the actual charge transport
and an effective current in proportion to $\nabla \times {\bf M}$
due to the magnetization $\bf M$ generated by the internal motion of the electrons.
In chiral $p$-wave superfluid the Cooper pair may be thought of as representing the atom 
and ${\bf l}$ the magnetic orbital moment $\bf M$.
Therefore the first term in the supercurrent describes the usual flow of Cooper pairs,
while the second one is an ``orbital" supercurrent driven by the spatial variation of the $l$-vector.

In this system, the spontaneous supercurrent is not a usual supercurrent $\mbox{\boldmath $\rho$}_s{\bf v}_s$ 
but an ``orbital" supercurrent ${\bf C} ( \nabla\times {\bf l})$ with the trap potential.
In the middle cross section in  Fig. \ref{f3}(a), the 
$l$-vector in-plane bending $(\nabla\times{\bf l})_z$ produces a perpendicular current $j_z$.
However, at the upper and lower planes in Fig. \ref{f3}(a), the supercurrent 
acquires the $j_x$ and $j_y$ components because of the non-vanishing $l_z$ component.
Therefore, the perpendicular current at the center bends such that 
the condensates are conserved.
It is clear from Fig. \ref{f3}(a) that the supercurrent circulates perpetually
along the $z$ direction parallel to the long axis of the trap. This result is non-trivial and
a remarkable manifestation in the topological nature of the texture.

In Fig. \ref{f3}(b) we also display the current patterns under rotation ($\Omega=0.3\omega_{\perp}$) for comparison
at rest. It is seen that the in-plane components
$j_x$ and $j_y$ increase, producing the in-plane circular current due to rotation, in particular 
in the middle cross section. Thus under rotations, the current consists of the spontaneous one along
the $z$ direction and the induced circular current in the plane.

\subsection{Half-quantum vortex under rotation}

\begin{figure}
\begin{center}
\includegraphics[width=12cm]{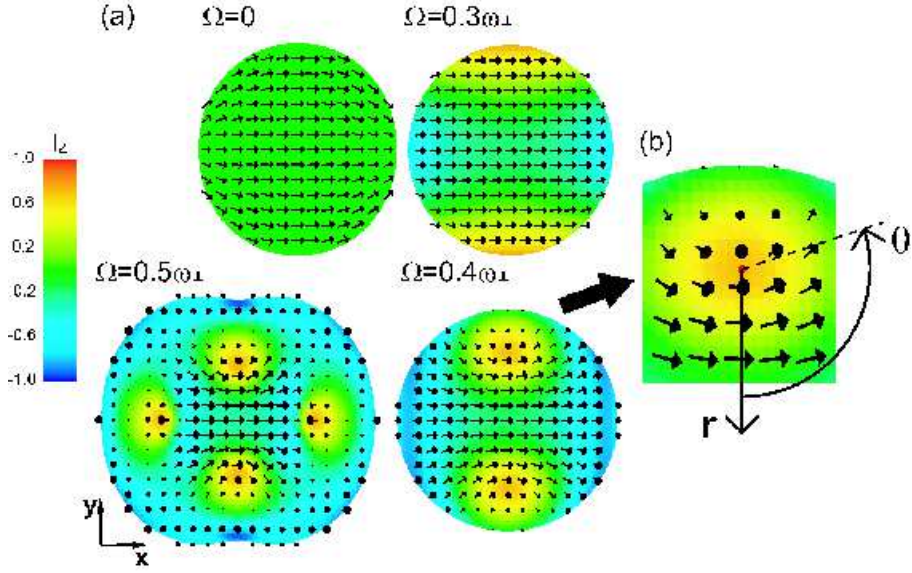}
\end{center}
\caption{(Color online) 
(a) Texture change with rotational speed $\Omega$ relative to the in-plane trap frequency $\omega_{\perp}$. 
Middle cross sections are displayed showing $l_z$  (color) and $l_x$ and $l_y$ components (arrows).
 As $\Omega$ increases, $l$-vectors acquire a negative $z$ component
(seen as a color change). At $\Omega=0.4\omega_{\perp}$ two plus HQVs enter from the $y$ direction
seen as yellow objects. 
At $\Omega=0.5\omega_{\perp}$ four plus HQVs and two minus HQVs are present and the condensates expand and deform.
(b) The enlarged figure of HQV where the local coordinates $(r, \theta)$ are shown.
}
\label{f4}
\end{figure}

\begin{figure}
\begin{center}
\includegraphics[width=8cm]{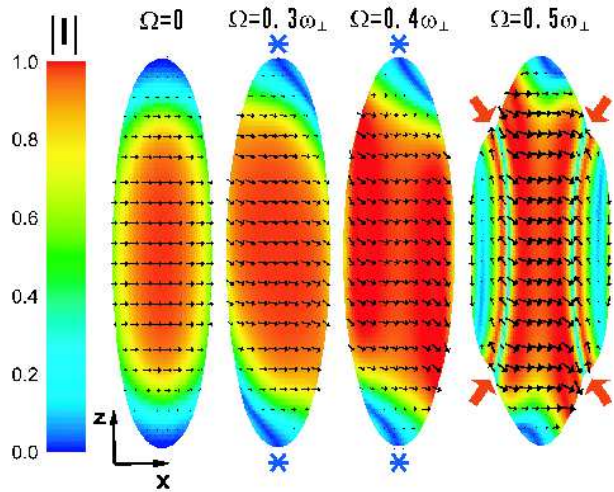}
\end{center}
\caption{(Color online) 
The $z$-$x$ cross sections corresponding to Fig. \ref{f4}.
The $l$-vectors lie almost in the $x$-$y$ plane at rest.
Under rotation two HQVs appear, shown as asterisks.
At $\Omega=0.5\omega_{\perp}$ two pairs of the HQVs
can be seen, indicated by red arrows.
}
\label{f6}
\end{figure}

In Fig. \ref{f4} where we depict the middle cross sections of the stable solutions 
for various rotation speeds (also see Fig. \ref{f6}). 
It is seen that under rotation, the 3D texture deforms continuously and smoothly.
As the rotational speed increases, 
the $l$-vectors in the $x$-$y$ plane  pointing in the $x$ direction
acquire a negative $z$ component, as seen by the color change from green to blue. 
This deformation yields in plane circular ``orbital" supercurrent as already shown in Fig. \ref{f3}(b).

Above a certain rotational speed ($\Omega=0.4\omega_{\perp}$), two HQVs
enter from the $y$ direction,
where the $l$-vector at the core pointing in the positive $z$ direction, as seen by the yellow objects.
The OP far away from the HQV core is described in terms of the local coordinated $(r, \theta)$
centered at the core (see Fig. \ref{f4}(b)) as 
\begin{eqnarray}
\Delta ( r \gg 1,\theta,\hat{\bf p}) = 
\exp[i(\theta /2 + \pi/2)] |A_{xy} | [ \sin(\theta /2) \hat{p}_x - \cos(\theta /2) \hat{p}_y] +  |A_z | \hat{p}_z,
 \label{hqv}
\end{eqnarray}
where $|A_{xy}|$ and $|A_z|$ are amplitude of the polar state 
characterized by the basis functions $\hat{p}_x$-$\hat{p}_y$  
($x$-$y$ polar state) and by $\hat{p}_z$ ($z$ polar state),  respectively.
The HQV is formed by only the $x$-$y$ polar state as is seen from \eqref{hqv}.
The above state \eqref{hqv} can be written as
\begin{equation}
\Delta ( r \gg 1,\theta,\hat{\bf p}) 
={1\over\sqrt{2}} |A_{xy} | (\hat{p}_+ +e^{i\theta } \hat{p}_-)+ |A_z|\hat{p}_0.
\end{equation}
The vortex core of this HQV has the orbital angular momentum with positive z component
because $\hat{p}_ +$ is non-vanishing there, so we define it as a plus HQV.
Similarly, one can construct a minus HQV with the identical vorticity 
but the local orbital angular momentum of the negative $z$ component in its core ($\hat{p}_ -$).

\begin{figure}
\begin{center}
\includegraphics[width=9cm]{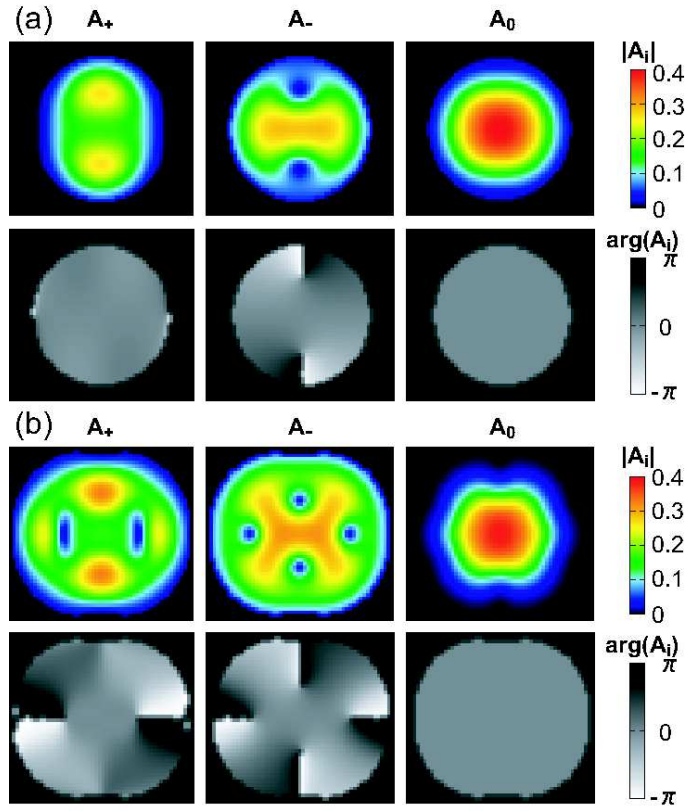}
\end{center}
\caption{(Color online) 
The amplitude (upper) and phase (lower) of 
the OP components  $A_+$, $A_-$ and $A_0$ in the $x$-$y$ plane at $z=0$ corresponding to Fig. \ref{f4}.
(a) At $\Omega=0.4\omega_{\perp}$ the vortices with winding number 1 enter $A_-$ component.
The depletion of the OP amplitude by the vortices is compensated by the growth of $A_+$ amplitude.
(b) At $\Omega=0.5\omega_{\perp}$
the vortices with winding number 1 appear.
The vortices on the $y$ and $x$ axis are plus and plus-minus HQVs, respectively.
}
\label{f5}
\end{figure}

The realized OP written as
\begin{eqnarray}
\Delta ({\bf r, \hat{p}}) = A_+({\bf r}) \hat{p}_+ + A_-({\bf r}) \hat{p}_- + A_0({\bf r}) \hat{p}_0,
\end{eqnarray}
\begin{eqnarray}
A_{\pm} = \mp {1\over\sqrt{2}} (A_x \mp iA_y), A_0=A_z
\end{eqnarray}
is shown in Fig. \ref{f5}(a) 
where the amplitudes (upper line) and phases (lower line) for each component $A_+, A_-, A_0$  are
displayed for $\Omega=0.4\omega_{\perp}$ corresponding to Fig. \ref{f4}.
It is seen that there are two vortices with the winding number 1 in $A_-$ component on the $y$ axis.
The depletion of the OP amplitude by the vortices is compensated by the growth of $A_+$ amplitude.
The HQV is embedded in the surrounding OP field.
As we walk around one of the HQV core, the $p$-wave pairing state changes in the following manner:
At $\theta =\pi$, the OP is described as $\Delta (\hat{\bf p}) = -|A_{xy} |\hat{p}_x + |A_z|\hat{p}_z$, 
namely this is the polar state.
At $\theta =0$, the OP is now $\Delta (\hat{\bf p}) = -i|A_{xy} |\hat{p}_y + |A_z|\hat{p}_z$.
This is a superposition of the polar state and the chiral state.
The spatial position where the arrow of $l$-vector near the HQV core vanishes is the polar state, 
and on the opposite side of HQV core, the chiral state is superposed by the polar state.
Since this HQV breaks the  reflection symmetry, the isolated HQV is energetically disfavored and
rather a pair of the HQV is advantageous. 
This is one of the reasons why we found a pair of the HQV at $\Omega=0.4\omega_{\perp}$.

Upon further increase in the rotational speed ($\Omega=0.5\omega_{\perp}$),
the HQVs enter further from the $x$ direction (see Fig. \ref{f4}).
They are different from above mentioned plus HQVs. As is seen from  Fig. \ref{f5}(b),
 a pair of plus-minus HQVs appears from the $x$ direction.
There are the vortices with the winding number 1 in each $A_+$ and $A_-$ component on the different position of the 
$x$ axis.
The depletion of the OP amplitude by the vortices is compensated 
by the growth of amplitude with the opposite orbital angular momentum component.
The minus HQV is situated more inside in the trap potential than plus HQV, 
because of the repulsive  interaction between the plus HQV on the $x$ axis and the $y$ axis.
It is an analogous situation for the HQVs in the $F=1$ spinor BEC\cite{Ji}.
We also notice from Fig. \ref{f4} ($\Omega=0.5\omega_{\perp}$) that  the condensate profile itself expands, deforms
and deviates clearly from a circular form due to  the non-trivial centrifugal energy.

Figure \ref{f6} shows a different view of Fig. \ref{f4}, displaying the $z$-$x$ cross section.
At rest, the $l$-vectors point almost to the $x$ direction.
As $\Omega$ increases, the downward $l_z$ component appears, 
which causes a counterclockwise circular ``orbital" supercurrent.
At the top and bottom ends of the system the HQVs appear as indicated by asterisks.
These blue lines show the polar state neighboring the HQV as mentioned above.
In the $\Omega=0.5\omega_{\perp}$ case, the side view of two pairs of plus-minus HQVs can be seen clearly.
The light blue lines indicated by the red arrows correspond to the polar states between plus and minus HQV.
These lines bend outward away from the center towards $\pm z$ direction.
Around the positions indicated by the red arrows the condensate profile is greatly deformed.
Because there the polar state dominates over the chiral state, which leads to the depletion of the condensate,
resulting in this deformation.

\section{Pancake Shape Trap}

In order to understand the geometric effect of the textures stabilized in cigar shape
trap, we examine the pancake trap case.
We consider the anisotropy $\lambda=3.0$  as an example, where we take $100\times100\times 80$ meshes 
with the cloud size $R_x=R_y=30$ and $R_z=10$ in the Thomas-Fermi approximation .
We fix the temperature at $t_x=t_y=0.5$ and $\alpha=0.95$.

\subsection{Axis symmetric texture at rest}

\begin{figure}
\begin{center}
\includegraphics[width=12cm]{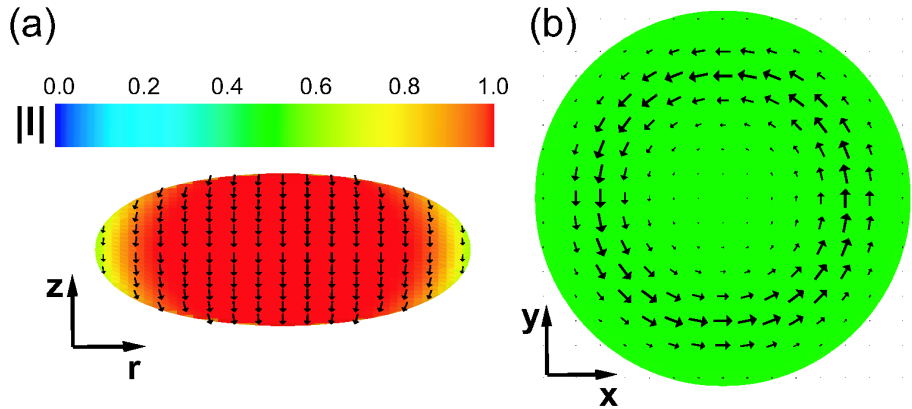}
\end{center}
\caption{(Color online) 
Stable texture in the pancake trap with $\lambda=3.0$ at rest.
(a) $l$-vector pattern in the $z$-$r$ plane. 
The pattern is axial symmetric around $z$.
(b) Spontaneous supercurrent in the $x$-$y$ plane at $z=0$.
}
\label{f7}
\end{figure}

Figure \ref{f7} displays the resulting $l$-vector texture (Fig. \ref{f7}(a))
and associated in-plane supercurrent (Fig. \ref{f7}(b)) at rest.
It is seen that most of the $l$-vectors point to the negative $z$ direction,
except for those near the upper and lower surface regions,
which acquire the  $r$-component of the radial direction.
The trap potential forces  the $l$-vector
to be parallel to the surface of condensates. This effect is especially strong when the
curvature of the surface is large.
Thus, in this case shown in Fig. \ref{f7}(a), the left and right ends of the system force the
$l$-vectors to point to the $z$ direction, giving an overall
$l$-vector configuration to the $z$ direction,
even for the vectors near the center.
The amplitude of the $l$-vectors decrease,
namely the polar state mixes with the chiral state towards the outside.

This stable texture is axis symmetric  around the $z$ axis,
which is different in the cigar shape trap.
This is because 
the effect of the trap potential is greater than one of the dipole-dipole interaction in the present situation ($\alpha=0.95$),
so that the chiral state consisting of the $\hat {p}_x$ and $\hat {p}_y$ components dominates over 
the polar state, resulting in the axis symmetric texture with respect to the $z$ axis.
If the influence of the dipole-dipole interaction is much greater, or $\alpha$ becomes small,
the stable texture is similar to those in the cigar shape trap.

The spatial variation of the amplitude of $l$-vectors towards the outer region
generates a circular ``orbital" supercurrent in the $x$-$y$ plane as shown in Fig. \ref{f7}(b).
Since the spontaneous supercurrent flows around the external rotational axis,
the axial symmetric texture is stable against low rotation.

\subsection{Vortices under rotation}

\begin{figure}
\begin{center}
\includegraphics[width=9cm]{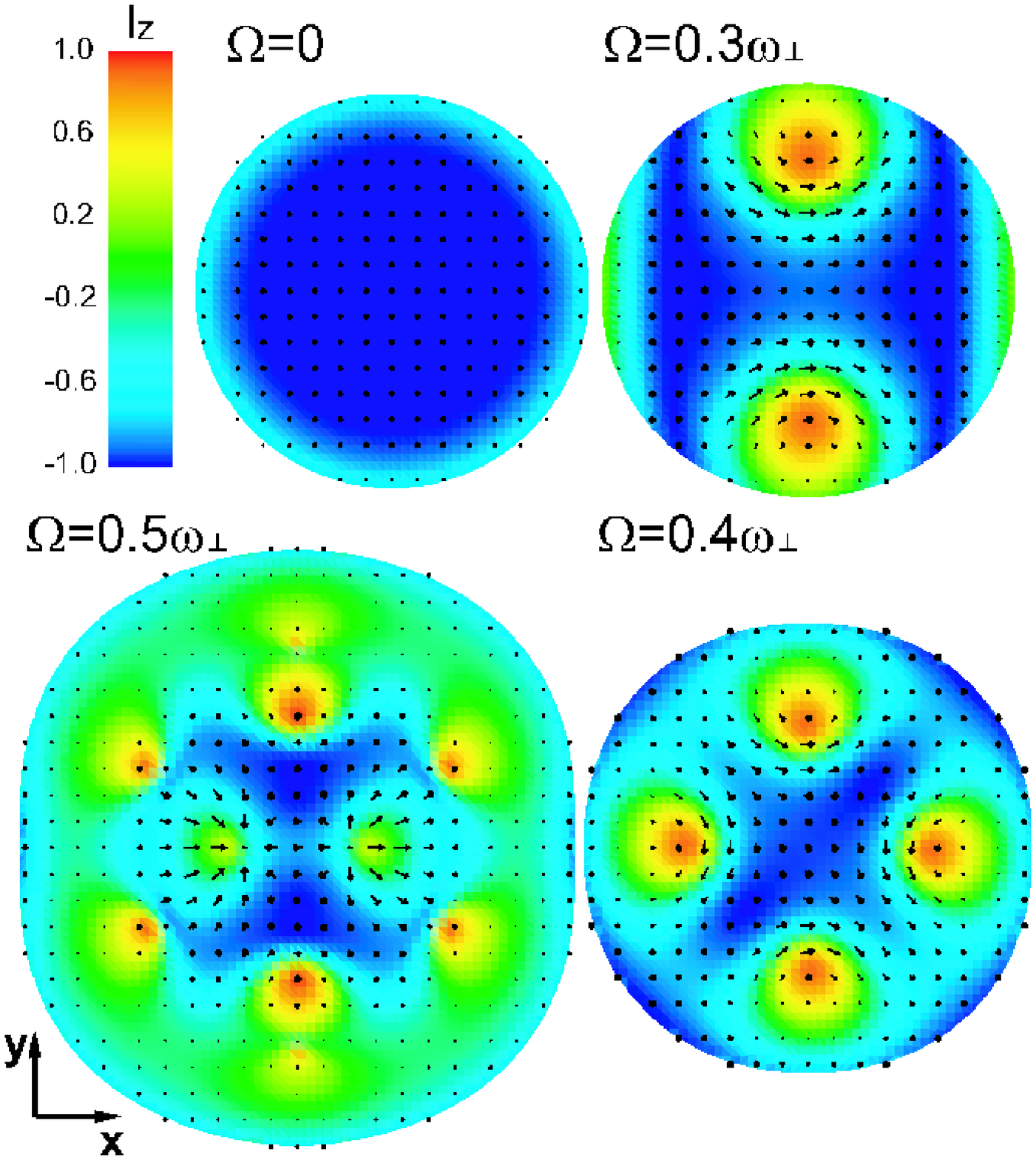}
\end{center}
\caption{(Color online) The $x$-$y$ cross sectional view of the textures
for the pancake shape trap with $\lambda=3.0$ for various rotations.
The two objects on the $y$ axis for $\Omega=0.3\omega_{\perp}$
are the HQVs. Compare those with Fig. \ref{f4} in cigar shape case.
}
\label{f8}
\end{figure}

\begin{figure}
\begin{center}
\includegraphics[width=8cm]{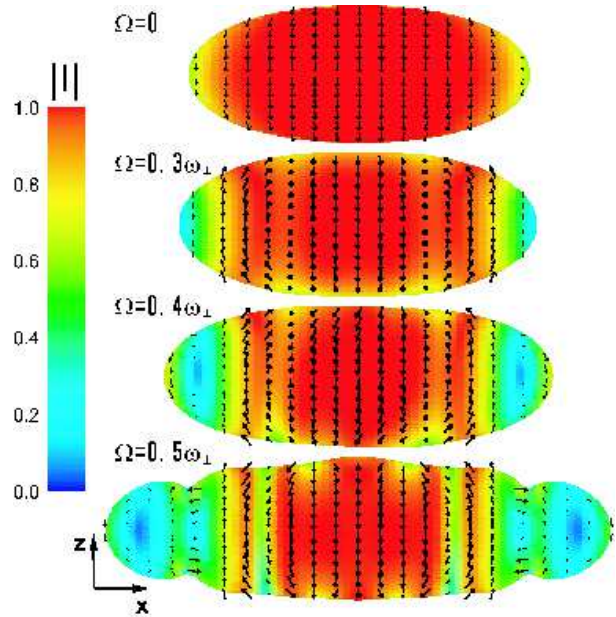}
\end{center}
\caption{(Color online) The $x$-$z$ cross sectional view of the same textures
in Fig. \ref{f8}. The polar state indicated by blue color gradually enter from the
right and left ends, making the system deform.
}
\label{f9}
\end{figure}

As shown in Fig. \ref{f8} where the $x$-$y$ cross section  at $z=0$ is displayed for
various rotations. Similarly to the previous texture changes in the cigar shape,
we see a pair of the HQV enter from the $y$ axis for $\Omega=0.3\omega_{\perp}$.
Those vortices are exactly the same HQV as in the cigar case (see Fig. \ref{f4}).
The following sequence upon increasing  $\Omega$  is very similar to the previous cigar case,
slightly differing its rotation speed where in the pancake case the sequence is shifted to lower
speed.
We illustrate the $x$-$z$ cross section at $y=0$ in Fig. \ref{f9}.
The almost $l$-vectors pointing to the negative $z$ direction at rest now deforms as $\Omega$
increases. From the right and left ends the polar state seen as blue color invades into the system.
At $\Omega=0.5\omega_{\perp}$ the condensate profile itself is deformed greatly
because of the dominance of the polar state associated with increasing HQVs.

\section{Summary and Discussions}

By minimizing the generic GL energy functional within weak coupling approximation,
we find stable textures for $p$-wave superfluids in neutral atomic gases confined  
in 3D harmonic trap potentials which are to be realized in near future.
In order to help detecting its superfluidity,
we examine two typical trap geometries, cigar and pancake shapes.
At rest, the obtained stable textures in both shapes exhibit the spontaneous supercurrent flow.
Under rotation, a pair of the HQVs enters in the condensates.
The isolated HQV is never stabilized in our calculations because of topological constrains.

It is interesting to note that
the direction of the spontaneously generated supercurrent 
at rest is always perpendicular to the
direction of the majority $l$-vectors, that is,
the condensates in the cigar and pancake case the $l$-vectors lie on 
the $x$-$y$ plane and point to the $z$ direction, respectively,
so the supercurrent flows toward the $z$ direction and in the $x$-$y$ plane.
This implies that the trap shape is critical in understanding 
and controlling the physics of the textures on the $p$-wave superfluid.

The $p$-wave Feshbach resonance occurs by using either the same species\cite{yip,leo}
or two species with different hyperfine states\cite{ho2,quader,ohashi}
whose difference in our context amounts to giving different values of the parameter $\alpha$
because the dipole-dipole interaction for a Cooper pair,
yielding the splitting between $m_l=\pm 1$ and $m_l=0$,
works differently for two cases.

As for Majorana particle generated in the vortex core in half-quantum vortex,
let us examine whether or not the present HQV accommodates
the Majorana particle in its core. The Mojorana condition is obviously fulfilled
when the half-quantum vortex is involved the $d$-vector, namely the spin
degrees of freedom\cite{kawakami}. However, in the present spinless HQV
it turns out by examining the Bogoliubov-de Gennes equation\cite{mizushima2} that 
the particles bound in the HQV core always have a finite energy, not at zero-energy,
which is indispensable for the Majorana condition.
Therefore, those are never the Majorana particles.
In order to find the Majorana particles, it is necessary to stabilize the
singular vortex with odd winding number in chiral superfluids\cite{mizushima2}.
In the present case, it might be realized for extremely flat pancake limit
where the $\hat{p}_z$ component becomes irrelevant.
The detailed energetics between the present HQV and the singular vortex
belongs to future study.

\section*{ACKNOWLEDGMENTS}

We wish to thank T. Ohmi, M. Ichioka, T. Mizushima, and T. Kawakami for useful discussions.

\appendix
\section{}
We start with the Hamiltonian in a co-rotating frame:
\begin{eqnarray}
H=H_0-{\bf \Omega \cdot L},
\end{eqnarray}
\begin{eqnarray}
H_0=\sum_i \left[ \left( p_i^2/2m \right) + U({\bf r}_i) \right] +V
\end{eqnarray}
where $H_0$ is the Hamiltonian in a non-rotating system, 
consisting of the kinetic energy term, the harmonic trap potential term $U$ and the interaction energy term $V$.
The angular velocity due to the external rotation is ${\bf \Omega}$ and
the angular momentum is ${\bf L}=\sum_i 1/2 \left( {\bf r}_i \times {\bf p}_i - {\bf p}_i \times {\bf r}_i \right)$.
We can write $H$ in the form
\begin{equation}
H = \sum_i \left[ \frac{1}{2m}\left( {\bf p}_i - m {\bf v}_{n,i} \right)^2 + U({\bf r}_i) \right] +V 
			 -\sum_i \frac{1}{2}m{\bf v}_{n,i}^2,
			 \label{Heff}
\end{equation}
where ${\bf v}_{n,i}={\bf \Omega \times r}_i$ is the ``normal fluid" velocity at the location of the particle $i$.
The last term is the centrifugal energy.

The gradient energy given by the first term of the Hamiltonian \eqref{Heff}
\begin{eqnarray}
f_{\rm grad}=\frac{7\zeta(3)N(0)}{16(\pi k_BT_c)^2}  
\left[\left( {\bf p}^* - m {\bf v}_{n} \right)_i A_j^*\right] \left[\left( {\bf p} - m {\bf v}_{n} \right)_k A_l\right]
\left\langle  v_{Fi} \hat{p}_j v_{Fk} \hat{p}_l \right\rangle_{\hat{\bf p}},
\end{eqnarray}
where $\langle \cdots \rangle_{\hat{\bf p}}$ denotes the Fermi surface average.
Taking the contraction of the subscripts in the Fermi surface average, the mean value is finite.
Replacing $\bf p$ with $-i\hbar \nabla$, we obtain the well-known form of the gradient energy\cite{wolfle}
\begin{equation}
f_{\rm grad}=K_1(\partial_i^* A_j^* )(\partial_i A_j )+K_2(\partial_i^* A_j^* )(\partial_j A_i )
+K_3(\partial_i^* A_i^* )(\partial_j A_j ) .
\end{equation}
In a similar manner, the centrifugal energy given by the last term of the Hamiltonian \eqref{Heff}
can be recast into
\begin{alignat}{2}
f_{\rm {cent}}=&-\frac{7\zeta(3)N(0)}{16(\pi k_BT_c)^2} 
[( m {\bf v}_{n} )_i A_j^*] [( m {\bf v}_{n} &&)_k A_l] 
\left\langle  v_{Fi} \hat{p}_j v_{Fk} \hat{p}_l \right\rangle_{\hat{\bf p}} \\
=&-\frac{m^2}{\hbar^2}[K_1({\bf \Omega \times r})_i A_j^*({\bf \Omega \times r})_i  A_j 
&&+K_2({\bf \Omega \times r})_i A_j^*({\bf \Omega \times r})_j  A_i \nonumber\\
&&&+K_3({\bf \Omega \times r})_i A_i^*({\bf \Omega \times r})_j  A_j ]. 
\end{alignat}
Taking ${\bf \Omega \parallel \hat{z}}$, and ${\bf \Omega \times r} = \Omega \rho \hat{\mbox{\boldmath $\theta$}}$,
we finally obtain the expression of  the centrifugal potential as
\begin{equation}
f_{\rm {cent}}=-{m^2\over \hbar^2}\Omega^2\rho^2(K_1A_i^* A_i + K_2 |A_{\theta}|^2 + K_3|A_{\theta}|^2 ).
\end{equation}

\appendix
\section{}

The bulk free energy \eqref{fbulk} is written as 
\begin{align}
f_{\rm{bulk}}=&-(1-t_x) ( |A_x|^2 + |A_y|^2) -(1-\alpha t_x)  |A_z|^2 \nonumber\\
&+(|A_x|^2 + |A_y|^2+ |A_z|^2)^2 +\frac{1}{2} (A_x^{*2} + A_y^{*2}+ A_z^{*2})(A_x^2 + A_y^2+ A_z^2). 
\end{align}
By taking $A_x$ real, $A_y=|A_y|e^{i\theta_y}$ and $A_z=|A_z|e^{i\theta_z}$ without 
loss of generality, we write it as 
\begin{align}
f_{\rm{bulk}}=& -(1-t_x) ( |A_x|^2 + |A_y|^2) -(1-\alpha t_x)  |A_z|^2 \nonumber\\
&+\frac{3}{2}(|A_x|^4 + |A_y|^4+ |A_z|^4) +2(|A_x|^2|A_y|^2+|A_y|^2|A_z|^2+|A_z|^2|A_x|^2) \nonumber\\
&+|A_x|^2|A_y|^2 \cos2\theta_y+ |A_y|^2|A_z|^2 \cos(2\theta_y-2\theta_z) +|A_z|^2|A_x|^2 \cos2\theta_z. 
\end{align}
There are six cases, which are possibly the minimum solutions:
(i) $\theta_y=\theta_z=0$, (ii) $\theta_y=0$, $\theta_z=\pi/2$, 
(iii) $\theta_y=\pi/2$, $\theta_z=0$, (iv)  $\theta_y=\pi/2$, $\theta_z=\pi/2$.
(v) $\theta_y=\pi/3$, $\theta_z=2\pi/3$ and (vi) $|A_x|^2=|A_y|^2=0$.

The relevant minimum solutions are found for (ii) and  (vi).
In the former case (ii) the solution is given by
\begin{eqnarray}
f_{\rm{bulk}}=-\frac{1}{16}[3(1-t_x)^2-2(1-t_x)(1-\alpha t_x)+3(1-\alpha t_x)^2 ] \nonumber
\end{eqnarray}
\begin{align}
|A_x|^2+|A_y|^2=&\frac{1}{8} [  3(1-t_x)-(1-\alpha t_x) ]\\
|A_z|^2=&\frac{1}{8} [ - (1-t_x)+3 (1-\alpha t_x) ] \nonumber
\end{align}
This solution is valid for
$2+(-3+\alpha)t_x\geq 0$, which determines the boundary between this phase called the A phase
and the single component phase called B phase below. 
The other solution for (vi) is expressed as 
\begin{align}
\begin{split}
f_{\rm{bulk}}=-\frac{1}{16}(1-\alpha t_x)^2 \\
|A_z|^2=\frac{1}{3} (1-\alpha t_x)
\end{split}
\end{align}
This B phase is described by a single component, thus 
it corresponds to the so-called polar phase.


\begin{thebibliography}{99}

\bibitem{dan}
D.M. Stamper-Kurn, M.R. Andrews,  A.P. Chikkatur, S. Inouye, H.J. Miesner,  J. Stenger, and W. Ketterle: 
Phys. Rev. Lett. {\bf 80}  (1998) 2027.

\bibitem{stenger}
J. Stenger, S. Inouye, D.M. Stamper-Kurn, H.J. Miesner, A.P. Chikkatur, and W. Ketterle: 
Nature {\bf 396} (1998) 345.

\bibitem{ohmi}
K. Machida and T. Ohmi: J. Phys. Soc. Jpn. {\bf 67} (1998) 1122.

\bibitem{ho}
T.-L. Ho: Phys. Rev. Lett. {\bf 81} (1998) 742.

\bibitem{leggett}
A.J. Leggett: Rev. Mod. Phys. {\bf 47} (1975)  331.

\bibitem{wolfle}
D. Vollhardt and P. W{\rm \"o}lfle: {\it The Superfluid phase of Helium 3}
(Taylor and Francis, London, 1990).

\bibitem{volovik}
G.E. Volovik: {\it Exotic Properties of Superfluid $^3$He}
(World Scientific, Singapore, 1992).

\bibitem{fetter}
A.L. Fetter: 
in {\it Progress in Low Temperature Physics}, ed D.F. Brewer (Elsevier Science Publishers,  Amsterdam, 1986) Vol. X, p. 1.  

\bibitem{salomaa}
M.M. Salomaa and G.E. Volovik: Rev. Mod. Phys. {\bf 59} (1987) 533.

\bibitem{machida}
K. Machida, M. Ozaki, and T. Ohmi: J. Phys. Soc. Jpn. {\bf 58} (1989) 4116;
K. Machida, T. Nishira, and T. Ohmi: J. Phys. Soc. Jpn. {\bf 68} (1999) 3364;
K. Machida and M. Ozaki: Phys. Rev. Lett. {\bf 66} (1991) 3293.

\bibitem{sauls}
J.A. Sauls: Adv. Phys. {\bf 43} (1994) 113.

\bibitem{buballa}
M. Buballa: Phys. Rep. {\bf 407} (2005) 205.

\bibitem{alford}
M.G. Alford, A. Schmitt, K. Rajagopal, and T. Sch{\rm \"a}fer: Rev. Mod. Phys. {\bf 80} (2008)  1455.


\bibitem{sarma}
See for example, C. Nayak, S.H. Simon, A. Stern, M. Freedman, and S. Das Sarma:
Rev. Mod. Phys. {\bf 80} (2008)  1083.

\bibitem{tsutsumi}
Y. Tsutsumi, T. Kawakami, T. Mizushima, M. Ichioka, and K. Machida:
Phys. Rev. Lett. {\bf 101} (2008)135302.


\bibitem{kawakami}
T. Kawakami, Y. Tsutsumi, and K. Machida:
Phys. Rev. B {\bf 79} (2009)  092506.

\bibitem{zheng}
J. Zheng, E.G.M. van Kempen, T. Bourdel, L. Khaykovich, J. Cubizolles, F. Chevy, M. Teichmann, L. Tarruell, 
S.J.J.M. F. Kokkelmans, and C. Salomon: Phys. Rev. A {\bf 70} (2004) 030702(R).

\bibitem{schunck}
C.H. Schunck, M.W. Zwierlein, C.A. Stan, S.M.F. Raupach, W. Ketterle, A.
Simoni, E. Tiesinga, C.J. Williams, and P.S. Julienne: Phys. Rev. A {\bf 71} (2005)  045601.

\bibitem{inada}
Y. Inada, M. Horikoshi, S. Nakajima, M. Kuwata-Gonogami, M. Ueda, and T. Mukaiyama:
Phys. Rev. Lett. {\bf 101} (2008)  100401.

\bibitem{jin1}
C.A. Regal, C. Ticknor, J.L. Bohn, and D.S. Jin: Nature {\bf 424}(2003) 47.

\bibitem{jin2}
C. Ticknor, C.A. Regal, D.S. Jin, and J.L. Bohn: Phys. Rev. A {\bf 69}  (2004)  042712.

\bibitem{jin3}
J.P. Gaebler, J.T. Stewart, J.L. Bohn, and D.S. Jin: Phys. Rev. Lett. {\bf 98}  (2007)  200403.

\bibitem{yip}
C.-H. Cheng and S.-K. Yip: Phys. Rev. Lett. {\bf 95} (2005)  070404.


\bibitem{leo}
V. Gurarie, L. Radzihovsky, and A.V. Andreev: Phys. Rev. Lett. {\bf 94} (2005)  230403.

\bibitem{ho2}
T.-L. Ho and R.B. Diener: Phys. Rev. Lett. {\bf 94} (2005)  090402.

\bibitem{quader}
K. Quader, R. Liao, and F. Popescu: Int. J. Mod. Phys. B {\bf 22} (2008)  4358.

\bibitem{ohashi}
Y. Ohashi: Phys. Rev. Lett. {\bf 94} (2005)  050403.

\bibitem{tsutsumi2}
Y. Tsutsumi and K. Machida: to be published in Phys. Rev. A.

\bibitem{baranov}
M.A. Baranov and D.S. Petrov: Phys. Rev. A {\bf 58}  (1998)  801(R).

\bibitem{mizushima}
T. Mizushima, N. Kobayashi, and K. Machida: Phys. Rev. A {\bf 70} (2004) 043613;
W.V. Pogosov, R. Kawate, T. Mizushima, and K. Machida: Phys. Rev. A {\bf 72} (2005) 063605.

\bibitem{Ji}
A.-C. Ji, W.M. Liu, J.L. Song, and F. Zhou: Phys. Rev. Lett. {\bf 101}  (2008)  010402.

\bibitem{mizushima2}
T. Mizushima, M. Ichioka, and K. Machida:  Phys. Rev. Lett. {\bf 101} (2008) 150409.

\end{thebibliography}
\end{document}